\documentclass[fleqn,reqno]{amsart}
\usepackage{amssymb,calrsfs}
\usepackage[mathscr]{eucal}
\allowdisplaybreaks[4]
\newtheorem{theorem}{Theorem}[section]
\newtheorem{lemma}[theorem]{Lemma}
\newtheorem{cor}[theorem]{Corollary}
\theoremstyle{definition}
\newtheorem{definition}{Definition}
\newtheorem{example}{Example}
\DeclareMathOperator{\inv}{inv}
\DeclareMathOperator{\diag}{diag}
\DeclareMathOperator{\Sym}{Sym}
\DeclareMathOperator{\Jac}{Jac}
\DeclareMathOperator{\id}{id}
\DeclareMathOperator{\odd}{od}
\DeclareMathOperator{\even}{ev}
\newcommand{\bwp}{\boldsymbol{\wp}}
\usepackage[backref]{hyperref}
\begin{document}
\title{Hyperelliptic Addition Law}
\author{V.M.~Buchstaber}
\address{RAS Steklov Mathematical institute,
8 Gubkina str., Moscow 117966, Russia}\email{buchstab@mendeleevo.ru}
\author{D.V.~Leykin}
\address{NASU Institute of Magnetism, 36-B Vernadsky str., Kiev 03142,
Ukraine}
\email{dile@imag.kiev.ua}
\date{2004 August 13}
\begin{abstract}
\noindent We construct an explicit form of the addition law for
hyperelliptic Abelian vector functions $\bwp$ and $\bwp'$. The
functions $\bwp$ and $\bwp'$ form a basis in the field of
hyperelliptic Abelian functions, i.e., any function from the field
can be expressed as a rational function of $\bwp$ and $\bwp'$.
\end{abstract}
\maketitle
\section{Introduction}
During the last 30 years the addition laws of elliptic functions
stay in the focus of the studies in the nonlinear equations of
Mathematical Physics. A large part of the interest was drawn to the
subject by the works of F.~Calogero~\cite{FC75,FC75a,FC76}, where
several important problems were reduced to the elliptic  addition
laws.  The term ``Calogero-Moser model'' being widely used in
literature, the papers caused a large series of publications, where
on one hand more advanced problems were posed and on the other hand
some advances were made in the theory of functional equations. The
``addition theorems'' for Weierstrass elliptic functions:
\begin{gather}
\frac{\sigma(u+v)\sigma(u-v)}{\sigma(u)^2\sigma(v)^2}=\wp(v)-\wp(u),\label{si}\\
(\zeta(u)+\zeta(v)+\zeta(w))^2=\wp(u)+\wp(v)+\wp(w),\quad
u+v+w=0,\label{ze}
\end{gather}
played the key r\^ole in the works of the that period. At the same
time, the development of the algebro-geometric methods of solution
of integrable systems \cite{dmn76,dkn01} employed in an essential
way the addition formulas for theta functions of several variables.
The same addition formulas were needed in applications of Hirota
method. In \cite{bk93,bk96} the addition theorems  for vector
Baker-Akhiezer functions of several variables are obtained and a
program is put forward to apply the addition theorems to problems of
the theory of integrable systems, in particular, to multidimensional
analogs of Calogero-Moser type systems.

The fundamental fact of the elliptic functions theory is that any
elliptic function can be represented as a rational function of
Weierstrass functions $\wp$ and $\wp'$. The corresponding result
(see \cite{bel97a,bel97b}) in the theory of hyperelliptic Abelian
functions is formulated as follows: any hyperelliptic function can
be represented as a rational function of vector functions
$\bwp=(\wp_{g1},\dots,\wp_{gg} )$ and
$\bwp'=(\wp_{gg1},\dots,\wp_{ggg})$, where $g$ is the genus of the
hyperelliptic curve on the Jacobi variety of which the field of
Abelian functions is built. In particular, for the genus $2$
sigma-function we obtain the following trilinear differential
addition theorem
\begin{equation*}
\big[2(\partial_{u_1}+\partial_{v_1}+\partial_{w_1})+
(\partial_{u_2}+\partial_{v_2}+\partial_{w_2})^3\big]\sigma(u)\sigma(v)\sigma(w)\big|_{u+v+w=0}=0.
\end{equation*}

In the present paper we find the explicit formulas for the addition
law of the vector functions $\bwp$ and $\bwp'$. As an application
the higher genus analogs of the Frobenius-Stickelberger formula
\eqref{ze} are obtained.

Our approach is based on the explicit construction of the groupoid
structure that is adequate to describe the algebraic structure of
the space of $g$-th symmetric powers of hyperelliptic curves.

\section{Algebraic groupoids}
\subsection{Topological groupoids}
\begin{definition} Take a topological space $\mathsf{Y}$.

A space $\mathsf{X}$ together with a mapping
$p_{\mathsf{X}}:\mathsf{X}\to Y$ is called \emph{a space over
$\mathsf{Y}$}. The mapping $p_{\mathsf{X}}$ is called ``an anchor''
in Differential Geometry.

Let two  spaces  $\mathsf{X}_1$ and $\mathsf{X}_2$ over $\mathsf{Y}$
be given. A mapping  $f:\mathsf{X}_1\to \mathsf{X}_2$ is called
\emph{a mapping over $\mathsf{Y}$}, if $p_{\mathsf{X}_2}\circ
f(x)=p_{\mathsf{X}_1}(x)$ for any point $x\in \mathsf{X}_1$.

By \emph{the direct product over $\mathsf{Y}$} of the spaces
 $\mathsf{X}_1$ and $\mathsf{X}_2$
 over $\mathsf{Y}$ we call the space $ \mathsf{X}_1\times_{\mathsf{Y}}\mathsf{X}_2=\{(x_1,x_2)\in
\mathsf{X}_1\times \mathsf{X}_2\mid
p_{\mathsf{X}_1}(x_1)=p_{\mathsf{X}_2}(x_2)\}$ together with the
mapping
$p_{\mathsf{X}_1\times_{\mathsf{Y}}\mathsf{X}_2}(x_1,x_2)=p_{\mathsf{X}_1}(x_1)$.
\end{definition}
The space  $\mathsf{Y}$ together with the identity mapping
$p_\mathsf{Y}$ is considered as the space over itself.

\begin{definition} A space $\mathsf{X}$ together with a mapping
$p_{\mathsf{X}}:\mathsf{X}\to \mathsf{Y}$ is called \emph{ a
groupoid over $\mathsf{Y}$}, if  there are defined the structure
mappings over $\mathsf{Y}$
\begin{equation*}
\mu: \mathsf{X}\times_\mathsf{Y} \mathsf{X}\to
\mathsf{X}\quad\text{and} \quad\inv: \mathsf{X}\to \mathsf{X}
\end{equation*}
that satisfy the axioms
\begin{enumerate}
\item $\mu(\mu(x_1,x_2),x_3)=
\mu(x_1,\mu(x_2,x_3)),$ provided $
p_{\mathsf{X}}(x_1)=p_{\mathsf{X}}(x_2)=p_{\mathsf{X}}(x_3).$
\item $\mu(\mu(x_1,x_2),\inv(x_2))=x_1$,  provided $
p_{\mathsf{X}}(x_1)=p_{\mathsf{X}}(x_2)$.
\end{enumerate}
The mapping  $\mu$ may not be defined for all pairs $x_1$ and $x_2$
from $\mathsf{X}$.
\end{definition}
\begin{definition} A groupoid structure on $\mathsf{X}$
over the space  $\mathsf{Y}$ is called commutative, if
 $\mu(x_1,x_2)=\mu(x_2,x_1)$, provided $
p_{\mathsf{X}}(x_1)=p_{\mathsf{X}}(x_2)$.
\end{definition}
\begin{definition}
A groupoid structure on the algebraic variety $\mathsf{X}$ over the
algebraic variety $\mathsf{Y}$ is called \emph{algebraic}, if the
mapping $p_\mathsf{X}$ as well as  the structure mappings
 $\mu$ and  $\inv$ are algebraic.
\end{definition}

\subsection{Algebraic groupoids related to plane curves}

We take as $\mathsf{Y}$ the space  $\mathbb{C}^{N}$ with coordinates
$\Lambda=(\lambda_{i})$, $i=1,\dots,N$. Let $f(x,y,\Lambda),$ where
$(x,y)\in\mathbb{C}^2$, be a polynomial in $x$ and $y$. Define the
 family of plane curves
\begin{equation*}
V=\{(x,y,\Lambda)\in\mathbb{C}^2\times\mathbb{C}^N\mid f(x,y,
\Lambda)=0\}.
\end{equation*}
We assume that at a generic value of $\Lambda$, genus of the curve
from $V$ has fixed value $g$.

Let us take as $\mathsf{X}$  the universal fiber-bundle of  $g$-th
symmetric powers of the algebraic curves from $V$. A point in
$\mathsf{X}$ is represented  by the collection of an unordered set
of $g$ pairs $(x_i,y_i)\in \mathbb{C}^2$ and  an $N$-dimensional
vector $\Lambda$ that are related by $f(x_i,y_i,\Lambda)=0$,
$i=1,\dots,g$.

The mapping $p_{\mathsf{X}}$ takes the collection to the point
$\Lambda\in \mathbb{C}^{N}.$

Let  $\phi(x,y)$ be an entire rational function on the curve  $V$
with the parameters $\Lambda$. A zero of the function $\phi(x,y)$ on
the curve $V$ is  the point  $(\xi,\eta)\in\mathbb{C}^2$, such that
$\{ f(\xi,\eta,\Lambda)=0,\phi(\xi,\eta)=0\}$. The total number of
zeros of the function $\phi(x,y)$ is called the order of
$\phi(x,y)$.

The further construction is based on the following fact.
\begin{lemma}Let  $\phi(x,y)$  be an order  $2g+k,$ $k\geqslant0$,
entire rational function on the curve
 $V$. Then the function  $\phi(x,y)$ is completely defined (up to a
constant with respect to $(x,y)$ factor) by any collection of  $g+k$
its zeros.
\end{lemma}
This fact is a consequence of Weierstrass gap theorem
(L{\"u}kensatz). In particular, an ordinary univariate polynomial is
an entire rational function on the curve of genus  $g=0$ and is
completely defined by the collection of all its zeros.

Let us construct the mapping $\inv$.

Let a point $U_{1}=\{[(x^{(1)}_i,y^{(1)}_i)],\Lambda\}\in
\mathsf{X}$ be given. Let $R^{(1)}_{2g}(x,y)$ be the entire rational
function
 of order $2g$ on the curve $V$ defined by the vector $\Lambda$,
 such that $R^{(1)}_{2g}(x,y)$  is zero in $U_1$, that is
$R^{(1)}_{2g}(x^{(1)}_i,y^{(1)}_i)=0$, $i=1,\dots,g$. Denote by
$[(x^{(2)}_i,y^{(2)}_i)]$ the complement of
$[(x^{(1)}_i,y^{(1)}_i)]$ in the set of zeros of
$R^{(1)}_{2g}(x,y)$. Denote by  $U_2$ the point in $\mathsf{X}$ thus
obtained and set $\inv(U_1)=U_2$.

So, the set of zeros of the function $R^{(1)}_{2g}(x,y)$, which
defines the mapping $\inv$, is the pair of points
$\{U_1,\inv(U_1)\}$  from $\mathsf{X}$ and
$p_{\mathsf{X}}(U_1)=p_{\mathsf{X}}(U_2)$.

\begin{lemma} The mapping $\inv$ is an involution, that is
$\inv\circ\inv(U_1)=U_1$.
\end{lemma}

Let us construct the mapping $\mu$.

Let two points $U_1=\{[(x^{(1)}_i,y^{(1)}_i)],\Lambda\}$ and
$U_2=\{[(x^{(2)}_i,y^{(2)}_i)],\Lambda\}$ from $\mathsf{X}$ be
given. Let $R^{(1,2)}_{3g}(x,y)$  be the entire rational function of
order $3g$ on the curve $V$ defined by the vector $\Lambda$, such
that $R^{(1,2)}_{3g}(x,y)$ is zero in  $\inv(U_1)$ and $\inv(U_2)$.
Denote by  $[(x^{(3)}_i,y^{(3)}_i)]$ the complementary $g$ zeros of
$R^{(1,2)}_{3g}(x,y)$ on the curve $V$. Denote by  $U_3$ the point
in $\mathsf{X}$ thus obtained and set $\mu(U_1,U_2)=U_3$.

So, the set of zeros of the function $R^{(1,2)}_{3g}(x,y)$, which
defines the mapping $\mu$, is the triple   of points
$\{\inv(U_1),\inv(U_2), \mu(U_1,U_2)\}$ from $\mathsf{X}$ and
$p_{\mathsf{X}}(U_1)=p_{\mathsf{X}}(U_2)=p_{\mathsf{X}}(\mu(U_1,U_2))$.
\begin{theorem}\label{universal-groupoid} The above mappings  $\mu$ and $\inv$ define the structure of
the commutative algebraic groupoid over $\mathsf{Y}=\mathbb{C}^{N}$
on the universal fiber-bundle $\mathsf{X}$ of
 $g$-th symmetric powers of the plane algebraic curves from the family
 $V$.
\end{theorem}
\begin{proof}
The mapping $\mu$ is symmetric with respect to $U_1$ and $U_2$, and
thus defines the commutative operation. By the construction the
mappings $p_\mathsf{X}$, $\mu$ and $\inv$ are algebraic.
\begin{lemma}\label{associatif} The mapping $\mu$ is associative.
\end{lemma}
\begin{proof}Let three points  $U_1, U_2$ and $U_3$ be given. Let us
assign
\begin{equation*}
U_{4}=\mu(U_1,U_2),\quad U_5=\mu(U_4,U_3),\quad
U_{6}=\mu(U_2,U_3),\quad U_7=\mu(U_1,U_6).\end{equation*} We have to
show that $U_5=U_7.$

Let $R_{3g}^{(i,j)}(x,y)$ be the function defining the point
$\mu(U_i,U_j)$, and  $R_{2g}^{(i)}(x,y)$ be the function defining
the point $\inv(U_i)$.

Consider the product $R_{3g}^{(1,2)}(x,y)R_{3g}^{(4,3)}(x,y)$. It is
a function of order  $6g$ with zeros at  \[\{\inv (U_1),\inv
(U_2),U_4,\inv(U_4),\inv(U_3),U_5\}.\]  Therefore, the function
\begin{equation*}
Q_1(x,y)=\frac{R_{3g}^{(1,2)}(x,y)R_{3g}^{(4,3)}(x,y)}{R_{2g}^{(4)}(x,y)}\end{equation*}
is an entire function of order  $4g$ with the zeros $\{\inv
(U_1),\inv (U_2),\inv(U_3),U_5\}$.

Similarly, the product $R_{3g}^{(2,3)}(x,y)R_{3g}^{(1,6)}(x,y)$ is a
function of order $6g$ with zeros at \[\{\inv (U_2),\inv
(U_3),U_6,\inv(U_6),\inv(U_1),U_7\}.\] Hence we find that
\begin{equation*}
Q_2(x,y)=\frac{R_{3g}^{(2,3)}(x,y)R_{3g}^{(1,6)}(x,y)}{R_{2g}^{(6)}(x,y)}\end{equation*}
is an entire function of order  $4g$ with the zeros $\{\inv
(U_1),\inv (U_2),\inv(U_3),U_7\}$.

The functions $Q_1(x,y)$ and  $Q_2(x,y)$ have order $4g$ and both
vanish at the points \[\{\inv (U_1),\inv (U_2),\inv(U_3)\}.\] Thus
by Weierstrass gap theorem $Q_1(x,y)=Q_2(x,y)$ and, therefore,
$U_5=U_7$.
\end{proof}

\begin{lemma}\label{weak_inverse}
The mappings  $\mu$ and $\inv$  satisfy the axiom \textup{2}.
\end{lemma}
\begin{proof}Let two points $U_1$ and $U_2$ be given. Assign
\begin{equation*}U_{3}=\mu(U_1,U_2),\quad U_4=\inv(U_2),\quad U_5=\mu(U_3,U_4).\end{equation*}
We have to show that  $U_5=U_1.$

Consider the product  $R_{3g}^{(1,2)}(x,y)R_{3g}^{(3,4)}(x,y)$,
which is the function of order $6g$ with the zeros \[\{\inv(U_1),
\inv(U_2), U_3,\inv(U_3),\inv(U_4),U_5\}.\] As
\[\inv(U_4)=\inv\circ\inv(U_2)=U_2,\] the function
\begin{equation*}
Q(x,y)=\frac{R_{3g}^{(1,2)}(x,y)R_{3g}^{(3,4)}(x,y)}{R_{2g}^{(2)}(x,y)R_{2g}^{(2)}(x,y)}
\end{equation*}
is the entire function of order $2g$ with zeros at
$\{\inv(U_1),U_5\}$, that is  $Q(x,y)=R_{2g}^{(5)}(x,y)$. Hence it
follows that $U_5=\inv\circ\inv(U_1)=U_1.$
\end{proof}
The Theorem is proved.
\end{proof}
The Lemma below is useful for constructing the addition laws on our
groupoids.
\begin{lemma}\label{even_part}
Given $U_1,U_2\in \mathsf{X}$, let us assign
 $U_{3}=\mu(U_1,U_2)$ and $U_{i+3}=\inv(U_i)$,
$i=1,2.$ Then
\begin{equation*}
R^{(3)}_{2g}(x,y)=
\frac{R^{(1,2)}_{3g}(x,y)R^{(4,5)}_{3g}(x,y)}{R^{(1)}_{2g}(x,y)R^{(2)}_{2g}(x,y)}.
\end{equation*}
\end{lemma}
The formula of Lemma \ref{even_part} is important because its left
hand side depends formally  on $U_3$ only, while the right hand side
is completely defined by the pair $U_1, U_2$.

The above general construction becomes  effective once we fix a
model of the family of curves, that is once the polynomial
$f(x,y,\Lambda)$ is given. We are especially interested in the
models of the form, cf. for instance \cite{bel99a,bl02,bl04},
\begin{equation*}
f(x,y,\Lambda)=y^n-x^s-\sum \lambda_{ns-in-js}x^i y^j,
\end{equation*}
where $\gcd(n,s)=1$ and the summation is carried out over the range
$0<i<s-1$, $0<j<n-1$ under the condition $ns-in-js\geqslant 0$. It
is important that a model of the kind (possibly with singular
points) exists for an arbitrary curve. At the generic values of
$\Lambda$ a curve in such a family has genus $g=(n-1)(s-1)/2.$ In
this paper we consider in detail the case $(n,s)=(2,2g+1)$, that is
the families of hyperelliptic curves.

\section{The structure of hyperelliptic groupoid on $\mathbb{C}^{3g}$}
A hyperelliptic curve  $V$ of genus $g$ is usually defined by a
polynomial of the form
\begin{equation*}
f(x,y,\lambda_{0},\lambda_{2},\dots)=y^2-4x^{2g+1}-\sum_{i=0}^{2g-1}\lambda_{i}x^i.
\end{equation*}
In this paper we apply the change of variables
\begin{equation*}(x,y,\lambda_{2g-1},\lambda_{2g-2},\dots,\lambda_{0})\to
(x, 2 y, 4\lambda_4,
4\lambda_6,\dots,4\lambda_{4g+2}),\end{equation*} in order to
simplify the formulas in the sequel. Below we study the
constructions related to the hyperelliptic curves defined by the
polynomials of the form
\begin{equation}
\label{hypp} f(x,y,\lambda_{4g+2},\lambda_{4g},\dots)=
y^2-x^{2g+1}-\sum_{i=0}^{2g-1}\lambda_{4g+2-2i}x^i.
\end{equation}
Let us introduce the grading by assigning $\deg x=2,$ $\deg y=2g+1$
and $\deg\lambda_k=k$. Then the polynomial $f(x,y,\Lambda)$ becomes
a homogeneous polynomial of the weight $4g+2.$

An entire function on  $V$ has  a unique representation as the
polynomial $R(x,y)=r_0(x)+r_1(x)y,$ where
$r_0(x),r_1(x)\in\mathbb{C}[x].$ In such representation we have not
more than one monomial $x^i y^j$ of each weight, by definition $\deg
x^i y^j=2i+j(2g+1)$. The order of a function $R(x,y)$ is equal to
the maximum of weights of the monomials that occur in $R(x,y)$. In
fact, on the set of zeros of the polynomial $R(x,y)$ we have
$y=-r_0(x)/r_1(x)$. Therefore, the zeros of $R(x,y)$ that lie on the
curve are defined by the roots  $x_1,\dots,x_m$ of the equation
$r_0(x)^2-r_1(x)^2(x^{2g+1}+\lambda_4 x^{2g-1}+\dots)=0.$ The total
number of the roots is equal  $m=\max(2\deg_x r_0(x),2g+1+2\deg_x
r_1(x))$, where $\deg_{x}r_j(x)$ denotes the degree of the
polynomial $r_j(x)$ in $x$, which is exactly the highest weight of
the monomials in $R(x,y)$.

In this case Weierstrass gap theorem  asserts that the sequence of
nonnegative integers $\{\deg x^i y^j\}$, $j=0,1$, $i=0,1,\dots$, in
ascending order has precisely  $g$ ``gaps'' in comparison to the
sequence of all nonnegative integers. All the gaps are less than
$2g$.
\begin{lemma}
For a given point $U_1=\{[(x^{(1)}_i,y^{(1)}_i)],\Lambda\}\in
\mathsf{X}$ the entire function  $R^{(1)}_{2g}(x,y)$ defining the
mapping  $\inv$ has the form
\begin{equation*}R^{(1)}_{2g}(x,y)=(x-x^{(1)}_1)\dotsc(x-x^{(1)}_g).\end{equation*}
\end{lemma}\begin{proof}
In fact, as $\deg y>2g$, any entire function of order $2g$ does not
depend on $y$.
\end{proof}
The function  $R^{(1)}_{2g}(x,y)$  defines the unique point
\begin{equation*}\inv(U_1)=\{[(x^{(1)}_i,-y^{(1)}_i)],\Lambda\},\end{equation*} which, obviously,
also belongs to  $\mathsf{X}$.

Let us construct the functions $R_{3g}^{(i,j)}(x,y)$ that have the
properties required in Lemmas \ref{associatif} and
\ref{weak_inverse}.
\begin{lemma}
Define the $(2g+1)$-dimensional row-vector
\begin{equation*}
m(x,y)=(1,x,\dots,x^{2g-1-\rho},y, y x,\dots, y x^{\rho}),\quad
\rho=\Big[\frac{g-1}{2}\Big],
\end{equation*}
which is  composed of all monomials $x^iy^j$ of weight not higher
than $3g$ (the restriction $j=0,1$ applies). Then, up to a factor
constant in $(x,y)$, the function $R_{3g}^{(1,2)}(x,y)$ is equal to
the determinant of the matrix composed of $2g+1$ rows $m(x,y)$,
 $m(x_i^{(1)},-y_i^{(1)})$ and
 $m(x_i^{(2)},-y_i^{(2)})$, $i=1,\dots, g.$
\end{lemma}\begin{proof}
By the construction the function
 $R_{3g}^{(1,2)}(x,y)$ vanishes at the points $\inv(U_1)$ and $\inv(U_2)$,
and is uniquely defined by this property. As
$\max(4g-2-2\rho,2g+1+2\rho)=3g$,  at fixed $\Lambda$ the function
$R_{3g}^{(1,2)}(x,y)$ has  $2g$  zeros at the given points of the
curve and the collection of zeros at $[(x^{(3)}_i,y^{(3)}_i)],$
$i=1,\dots,g$, which defines the unique point $U_3$ in $\mathsf{X}$.
\end{proof}

Let $\Sym^n(\mathsf{M})$ denote the  $n$-th symmetric power of the
space $\mathsf{M}$. A point of the space $\Sym^n(\mathsf{M})$ is an
unordered collection $[m_1,\dots,m_n],$ $m_i\in \mathsf{M}$.

Consider the space $\mathsf{S}=\Sym^{g}(\mathbb{C}^2)\times
\mathbb{C}^g$ and let us define the mapping
$p_{\mathsf{S}}:\mathsf{S}\to \mathsf{Y}=\mathbb{C}^{2g}.$ Take a
point $T=\{[(\xi_i,\eta_i)],Z\}\in \mathsf{S}$. Denote by
$\mathcal{V}$ the Vandermonde matrix, composed of $g$ rows $(1,
\xi_{i},\dots,\xi_{i}^{g-1})$, denote by $\mathcal{X}$ the diagonal
matrix $\diag (\xi_1^g,\dots,\xi_g^g)$ and by  $\mathcal{Y}$ the
vector $(\eta_1^2-\xi_1^{2g+1},\dots,\eta_g^2-\xi_g^{2g+1})^t$. Set
\begin{equation*}p_S(T)=(Z_1,Z_2),\quad\text{where}\quad
Z_1=\mathcal{V}^{-1}\mathcal{Y}-(\mathcal{V}^{-1}\mathcal{X}\mathcal{V})
Z  \quad\text{and}\quad   Z_2=Z.\end{equation*} It is clear that the
domain of definition of the mapping $p_\mathsf{S}$ is the open and
everywhere dense subset  $\mathsf{S}_0$ in $\mathsf{S}$ consisting
of the points $\{[(\xi_i,\eta_i)],Z\}$ such that the determinant
$\mathcal{V}$ does not vanish.

We define the mappings  $\gamma:\mathsf{X}\to \mathsf{S}$ and
$\delta: \mathsf{S}\to \mathsf{X}$ by the following formulas: let
$U\in \mathsf{X}$ and $T\in \mathsf{S}$, then
\[
\gamma(U)=\gamma(\{[(x_i,y_i)],\Lambda\})=\{[(x_i,y_i)],\Lambda_2\},
\]
where $ \Lambda_2=(\lambda_{2(g-i)+4}),$ $i=1,\dots,g,$
\[
\delta(T)=\delta(\{[(\xi_i,\eta_i)],Z\})=\{[(\xi_i,\eta_i)],P_S(T)\}.
\]
By the construction, the mappings are mappings over $\mathsf{Y}$.
The domain of definition of $\delta$ coincides with the domain
$\mathsf{S}_0\subset \mathsf{S}$  of definition of the mapping
$p_\mathsf{S}$. Let $T\in \mathsf{S}_0$, then $\gamma\circ
\delta(T)=T$. Let $\gamma(U)\in \mathsf{S}_0$, then $\delta\circ
\gamma(U)=U$. Thus we have
\begin{lemma} \label{birazio} The mappings $\gamma$ and
$\delta$ establish the birational equivalence of the spaces
$\mathsf{X}$ and $\mathsf{S}$ over $\mathsf{Y}=\mathbb{C}^{2g}.$
\end{lemma}
The assertion of Lemma \ref{birazio} helps to transfer onto the
space $\mathsf{S}$ the groupoid over $\mathsf{Y}$ structure, which
is introduced by Theorem \ref{universal-groupoid} on the space
$\mathsf{X}$. Let $T_1, T_2\in \mathsf{S}$. The birational
equivalence induces the mappings  $\mu_{*}$ and $\inv_{*}$ that are
defined by the formulas
\begin{equation*}
\mu_{*}(T_1,T_2)=\gamma\circ\mu(\delta(T_1),\delta(T_2)),\quad
\inv_{*}(T_1)=\gamma\circ\inv\circ\,\delta(T_1).
\end{equation*}

\begin{theorem}
The mappings  $\mu_{*}$ and $\inv_{*}$ define the structure of
commutative algebraic groupoid over the space
$\mathsf{Y}=\mathbb{C}^{2g}$ on the space $\mathsf{S}$.
\end{theorem}

Let us proceed to constructing the structure of algebraic groupoid
over  $\mathbb{C}^{2g}$ on the space $\mathbb{C}^{3g}$. The
classical Vi{\`e}te mapping is the homeomorphism of spaces
$\Sym^g(\mathbb{C})\to \mathbb{C}^g$. Let us use Vi{\`e}te mapping
to construct a birational equivalence
$\varphi:\Sym^{g}(\mathbb{C}^2)\to \mathbb{C}^{2g}$.

Let $[(\xi_j,\eta_j)]\in \Sym^{g}(\mathbb{C}^2)$ and
$(p_{2g+1},p_{2g},\dots,p_2)\in \mathbb{C}^{2g}$. Let us assign
$P_{\odd}=(p_{2g+1},p_{2g-1},\dots,p_3)^{t}$,
$P_{\even}=(p_{2g},p_{2g-2},\dots,p_2)^{t}$ and
$X=(1,x,\dots,x^{g-1})^t$. Define the mapping  $\varphi$ and its
inverse $\psi$ with the help of the relations
\begin{gather*}
x^g-\sum_{i=1}^g p_{2i}x^{g-i}=x^g-X^tP_{\even}=\prod_{j=1}^g(x-\xi_j),\\
\eta_i=P_{\odd}^t X|_{x=\xi_i},\quad i=1,\dots,g.
\end{gather*}
Note, that $\varphi$ is a rational mapping, while $\psi$ is a
nonsingular algebraic mapping.

Using the mapping $\varphi$, we obtain the mapping
\begin{equation*}\varphi_1=\varphi\times\id:\mathsf{S}=
\Sym^g(\mathbb{C}^2)\times\mathbb{C}^g\to\mathbb{C}^{2g}\times\mathbb{C}^g\cong
\mathbb{C}^{3g}\end{equation*} and its inverse
$\psi_1=\psi\times\id$.

\emph{The companion matrix  of a polynomial $x^g-X^tP_{\even}$} is
the matrix
\begin{equation*}C=\sum_{i=1}^ge_i(e_{i-1}+p_{2(g-i+1)}e_{g})^t,\end{equation*}
where $e_i$ is the $i$-th basis vector in $\mathbb{C}^g$. Its
characteristic polynomial $|x\cdot 1_g-C|$ is $x^g-X^tP_{\even}.$

\begin{example}The companion matrices $C$ of the polynomials$x^g-X^tP_{\even}$
for $g=1,2,3,4$, have the form
\begin{equation*}
p_2,\quad
\begin{pmatrix}
0&p_4\\
1&p_2
\end{pmatrix},
\quad
\begin{pmatrix}
0&0&p_6\\
1&0&p_4\\
0&1&p_2
\end{pmatrix}
,\quad
\begin{pmatrix}
0&0&0&p_8\\
1&0&0&p_6\\
0&1&0&p_4\\
0&0&1&p_2
\end{pmatrix}.
\end{equation*}
Note, that the companion matrix for $g=k$ is included in the
companion matrix for $g>k$ as the lower right $k\times k$ submatrix.
\end{example}
We make use of the following property of a companion matrix.
\begin{lemma} Let the polynomial
$p(x)=x^g-\sum_{i=1}^g p_{2i}x^{g-i}$ and one of its roots  $\xi$ be
given. Set
 $\Upsilon=(1,\xi,\dots,\xi^{g-1})^t$. Then
the relations
\begin{equation*}
\xi^k \Upsilon^tA=\Upsilon^tC^kA,\quad k=0,1,2,\dots,
\end{equation*}
hold for an arbitrary vector $A\in \mathbb{C}^g$. \label{companion}
\end{lemma}
\begin{proof}Let $A=(a_1,\dots,a_g),$ then
\begin{gather*}
\xi \Upsilon^t A=\xi\sum_{i=1}^g
a_i\xi^{i-1}=\\
\qquad a_g\xi^g+\sum_{i=2}^g a_{i-1}\xi^{i-1}
=\sum_{i=1}^g((1-\delta_{i,1})a_{i-1}+a_gp_{2(g-i+1)})\xi^{i-1}.
\end{gather*}
Thus the Lemma holds for $k=1$. One can complete the proof by
induction.
\end{proof}
The mapping $p_{\mathbb{C}^{3g}}:\mathbb{C}^{3g}\to\mathbb{C}^{2g},$
with respect to which $\varphi_1$ is a mapping over
$\mathbb{C}^{2g}$, is given by the formula
\begin{gather}\label{p_C^3g}
p_{\mathbb{C}^{3g}}(P,Z)=(Z_1,Z_2),\\
Z_1=\Big(\sum_{i=1}^g p_{2i+1}C^{g-i}\Big)P_{\odd}-C^{g}(C
P_{\even}+Z),\quad Z_2=Z. \notag
\end{gather}
Using Lemma \ref{companion} one can directly verify the ``over''
property, that is, that
$p_{\mathbb{C}^{3g}}\circ\varphi_1(T)=p_\mathsf{S}(T)$ for any $T\in
\mathsf{S}_0$.

Let $A_1, A_2\in \mathbb{C}^{3g}$. The birational equivalence
 $\varphi_1$ induces the mappings
$\mu_{**}$ and $\inv_{**}$ defined by formulas
\begin{equation*}
\mu_{**}(A_1,A_2)=\varphi_1\circ\mu_{*}(\psi_1(A_1),\psi_1(A_2)),\quad
\inv_{**}(A_1)=\varphi_1\circ\inv_{*}\circ\,\psi_1(A_1).
\end{equation*}
\begin{theorem}
The mappings  $\mu_{**}$ and $\inv_{**}$  define the structure of
commutative algebraic groupoid over the space
$\mathsf{Y}=\mathbb{C}^{2g}$ on the space $\mathbb{C}^{3g}$.
\end{theorem}
\section{The addition law of the hyperelliptic groupoid on $\mathbb{C}^{3g}$}
In what follows we use the shorthand notation
\begin{equation*}
\overline{A}=\inv_{**}(A)\quad\text{and}\quad A_1\star
A_2=\mu_{**}(A_1,A_2).
\end{equation*}
\begin{lemma}
Let $A=(P_{\even},P_{\odd},Z)\in \mathbb{C}^{3g}.$ Then
\begin{equation*}
\overline{A}=(P_{\even},-P_{\odd},Z).
\end{equation*}
\end{lemma}
Introduce the $(g\times \infty)$-matrix
\begin{equation*}
K(A)=\big(Y,C Y, C^2Y,\dots\big),
\end{equation*}
that is composed of the $(g\times 2)$-matrix
$Y=(P_{\even},P_{\odd})$ with the help of the companion matrix $C$
of the polynomial$x^g-X^tP_{\even}$. Denote by $L(A)$ the matrix
composed of the first  $g$ columns of $K(A)$, and denote by
$\ell(A)$ the  $(g+1)$-st column of $K(A)$.
\begin{theorem} Let $A_1,A_2\in \mathbb{C}^{3g}$ and let
$A_3={A}_1\star A_2$, then
\begin{equation*}
\mathrm{rank}
\begin{pmatrix}
1_g&L(\overline{A}_1)&\ell(\overline{A}_1)\\
1_g&L(\overline{A}_2)&\ell(\overline{A}_2)\\
1_g&L(A_3)&\ell(A_3)\\
\end{pmatrix}<2g+1.
\end{equation*}\label{additto_1}
\end{theorem}
\begin{proof} Suppose  the points  $U_i=\delta\circ\psi_1(A_i),$
$i=1,2$, are defined. We rewrite the function $R^{(1,2)}_{3g}(x,y)$
as a linear combination of  monomials
\begin{equation*}
R^{(1,2)}_{3g}(x,y)=\sum_{i,j,w(i,j)\geqslant 0} h_{w(i,j)}x^iy^j
=r_1(x)y+x^gr_2(x)+r_3(x),\end{equation*} where
$w(i,j)=3g-(2g+1)j-2i$,
\begin{equation*}\begin{gathered}
r_1(x)=\sum_{i=0}^{\rho}h_{g-2i-1}x^i,\\
r_2(x)=\sum_{i=0}^{g-\rho-1}h_{g-2i}x^i,\\
r_3(x)=\sum_{i=0}^{g-1}h_{3g-2i}x^i,
\end{gathered} \qquad
\rho=\Big[\frac{g-1}{2}\Big].
\end{equation*}
Let us set $h_0=1.$

We assign weights to the parameters $h_k$ by the formula $\deg
h_k=k$. Then $\deg R^{(1,2)}_{3g}(x,y)=3g$.

Let $A=(P_{\even},P_{\odd},Z)\in\mathbb{C}^{3g}$ be the point
defining  any of the collections of $g$ zeros of the function
$R_{3g}^{(1,2)}(x,y).$ Consider the function
$Q(x)=R^{(1,2)}_{3g}(x,X^tP_{\odd})$. By the construction
$Q(\xi)=0,$ if $(x^g-X^tP_{\even})|_{x=\xi}=0.$ Let us apply Lemma
\ref{companion}. We obtain
\begin{equation}
Q(\xi)= \Upsilon^t\big(
r_1(C)P_{\odd}+r_2(C)P_{\even}+H_1\big),\label{quan}
\end{equation}
where $H_{1}=(h_{3g},h_{3g-2},\dots, h_{g+2})$. Using the above
notation, we come to the relation
\begin{equation*}
Q(\xi)=\Upsilon^t\big( H_1+L(A)H_2+\ell(A)\big),
\end{equation*}
where $H_2=(h_g,h_{g-1},\dots,h_1)$. Suppose the polynomial
$x^g-X^tP_{\even}$ has no multiple roots, then from the equalities
$Q(\xi_j)=0,$ $j=1,\dots,g$ one can conclude that
\begin{equation}
H_1+L(A)H_2+\ell(A)=0. \label{juan}\end{equation}

By substituting the points $\overline{A}_1, \overline{A}_2$ and
$A_3$ to \eqref{juan}, we obtain the system of $3g$ linear
equations, which is satisfied by the coefficients $H_1,H_2$ of the
entire function $R^{(1,2)}_{3g}(x,y)$. The assertion of the Theorem
is the compatibility condition of the system of linear equations
obtained.
\end{proof}
\begin{cor}The vectors $H_1,H_2$ of coefficients of the entire function
 $R^{(1,2)}_{3g}(x,y)$ are expressed by the formulas
\begin{gather*}
H_2(A_1,A_2)=-\big[L(\overline{A}_1)-
L(\overline{A}_2)\big]^{-1}(\ell(\overline{A}_1)-\ell(\overline{A}_2)),\\
H_1(A_1,A_2)=-\frac{1}{2}\big[(\ell(\overline{A}_1)+
\ell(\overline{A}_2))-(L(\overline{A}_1)+L(\overline{A}_2))H_2(A_1,A_2)\big].
\end{gather*}
as vector functions of the points $A_1$ and $A_2$
 from $\mathbb{C}^{3g}$.
\end{cor}
Now, we know the coefficients $H_1,H_2$ of $R^{(1,2)}_{3g}(x,y)$ and
we can give the expression of $P^{(3)}_{\odd}$ as a function of
$A_1, A_2$ and $P^{(3)}_{\even}$. It follows from \eqref{quan} that
the following assertion holds.
\begin{lemma}
\begin{equation}\label{p3_odd}
P_{\odd}^{(3)}=-\big[r_1(C^{(3)})\big]^{-1}(H_1+r_2(C^{(3)})P_{\even}^{(3)}),
\end{equation}
where  $C^{(3)}$ is the companion matrix of the polynomial
$x^g-X^tP_{\even}^{(3)}$.
\end{lemma}
Let us find the explicit formula for the function
$R_{3g}^{(1,2)}(x,y)$ as a function of the points $A_1$ and $A_2$.
We introduce the $((2g+1)\times\infty)$-matrix
\begin{equation*}
F(x,y;A_1,A_2)=
\begin{pmatrix}
X^t&\mathcal{K}(x,y)\\
1_g&K(A_1)\\
1_g&K(A_2)
\end{pmatrix},
\end{equation*}
where $\mathcal{K}(x,y)=(x^g,y,\dots,x^{g+k},y x^k,\dots).$

Denote by $G(x,y;A_1,A_2)$ the matrix composed of the first $2g+1$
columns of the matrix $F(x,y;A_1,A_2).$
\begin{theorem}The entire rational function
$R_{3g}^{(1,2)}(x,y)$ defining the operation $A_1\star A_2$ has the
form
\begin{equation}\label{R_explicit}
R_{3g}^{(1,2)}(x,y)=\frac{|G(x,y;\overline{A}_1,
\overline{A}_2)|}{|L(\overline{A}_2)-L(\overline{A}_1)|}.
\end{equation}
\end{theorem}
By a use of the formula \eqref{R_explicit} and Lemma \ref{even_part}
we can find  $P_{\even}^{(3)}.$ Similar to the condition of Lemma
\ref{even_part} denote $A_4=\overline{A}_1$ and
$A_5=\overline{A}_2$. One can easily show that
$R_{3g}^{(4,5)}(x,y)=R_{3g}^{(1,2)}(x,-y)$. Thus, the product
 $R_{3g}^{(1,2)}(x,y)\,R_{3g}^{(4,5)}(x,y)$ is an even function in
 $y$. Set
\begin{gather*}
\Phi(x,y^2)=R_{3g}^{(1,2)}(x,y)\,R_{3g}^{(4,5)}(x,y)\\ \qquad=
(-1)^g\frac{|G(x,y;A_1,A_2)|}{|L(A_2)-L(A_1)|}\frac{|G(x,-y;A_1,A_2)|}{|L(A_2)-L(A_1)|}.
\end{gather*}
Therefore, $\Phi(x,y^2)$, as a function on the curve $V$, is the
polynomial in  $x$ and the parameters $Z_1$ and $Z_2$. The values of
$Z_1$ and $Z_2$ are defined by the mapping $p_{\mathbb{C}^{3g}}$
according to \eqref{p_C^3g}, and
$p_{\mathbb{C}^{3g}}(A_1)=p_{\mathbb{C}^{3g}}(A_2)=(Z_1,Z_2)$.
Namely, we have
\begin{equation*}
R_{3g}^{(1,2)}(x,y)\,R_{3g}^{(4,5)}(x,y)=
\Phi(x,x^{2g+1}+x^g X^tZ_2+X^tZ_1).\end{equation*}
Lemma \ref{even_part} asserts that dividing the polynomial
$\Phi(x,x^{2g+1}+x^g X^tZ_2+X^tZ_1)$ by
$(x^g-X^tP^{(1)}_{\even})(x^g-X^tP^{(2)}_{\even})$ gives the zero
remainder and the quotient equal $x^g-X^tP^{(3)}_{\even}$. Thus, the
calculation is reduced to the classical algorithm of polynomial
division.
\begin{theorem}\label{zakon} Consider the space
$\mathbb{C}^{3g}$ together with the mapping $p_{\mathbb{C}^{3g}}$
defined by \eqref{p_C^3g} as a groupoid over $\mathbb{C}^{2g}$. Let
$A_1=(P^{(1)}_{\even},P^{(1)}_{\odd}, Z)$ and
$A_2=(P^{(2)}_{\even},P^{(2)}_{\odd}, Z)$ be the points from
$\mathbb{C}^{3g}$ such that
$p_{\mathbb{C}^{3g}}(A_1)=p_{\mathbb{C}^{3g}}(A_2)=(Z_1,Z_2)\in\mathbb{C}^{2g}$.

Then the addition law has the form $ A_1\star A_2=A_3,$ where the
coordinates of the point $A_3=(P^{(3)}_{\even},P^{(3)}_{\odd}, Z)$
are given by the formulas
\begin{gather*}x^g-X^tP_{\even}^{(3)}= \frac{\Phi(x,x^{2g+1}+x^g
X^tZ_2+X^tZ_1)}{(x^g-X^tP^{(1)}_{\even})(x^g-X^tP^{(2)}_{\even})},
\\
P_{\odd}^{(3)}=-\big[r_1(C^{(3)})\big]^{-1}(H_1+r_2(C^{(3)})P_{\even}^{(3)}).
\end{gather*}
\end{theorem}
\begin{example}\label{genus1.1}
Let $g=1.$ The family of curves $V$ is defined by the polynomial
\begin{equation*}
f(x,y,\Lambda)=y^2-x^3-\lambda_4x-\lambda_6.
\end{equation*} In the coordinates $(\lambda_6,\lambda_4)$ on
$\mathbb{C}^2$ and $(p_2,p_3,z_4)$ and $\mathbb{C}^3$ the mapping
 $p_{\mathbb{C}^3}$ is given by the formula
\begin{equation*}
(\lambda_6,\lambda_4)=(p_3^2-p_2(p_2^2+z_4),z_4)
\end{equation*}
Let us write down the addition formulas for the points on the
groupoid $\mathbb{C}^3$ over $\mathbb{C}^2$. Set
$A_1=(u_2,u_3,\lambda_4)$, $A_2=(v_2,v_3,\lambda_4)$ and suppose
$p_{\mathbb{C}^3}(A_1)=p_{\mathbb{C}^3}(A_2)=(\lambda_6,\lambda_4)$.

Let $A_1\star A_2=A_3=(w_2,w_3,\lambda_4)$.

We have: $R_2^{(1)}(x,y)=x-u_2,$ $
 L(A_1)=u_2$, $\ell(A_1)=u_3,$ and so on.
\begin{equation*}F(x,y,A_1,A_2)=
\begin{pmatrix}
1&x&y&x^2&yx&\dots\\
1&u_2&u_3&u_2^2&\hdotsfor{2}\\
1&v_2&v_3&\hdotsfor{3}
\end{pmatrix}.
\end{equation*}
Thus, the function defining the operation  $A_1\star A_2$ has the
expression
\begin{equation*}
R_{3}^{(1,2)}(x,y)=y+\frac{v_3-u_3}{v_2-u_2}x-\frac{u_2v_3-u_3v_2}{v_2-u_2}.
\end{equation*}
Whence, we find $r_1(x)=1,$ $r_2(x)=\dfrac{v_3-u_3}{v_2-u_2},$
$H_1=-\dfrac{u_2v_3-u_3v_2}{v_2-u_2}$ and, by \eqref{p3_odd},
\begin{equation*}
w_3=\frac{u_2v_3-u_3v_2}{v_2-u_2}-\frac{v_3-u_3}{v_2-u_2}w_2.
\end{equation*}
Further, $\Phi(x, x^3+\lambda_4x+\lambda_6)=x^3+\lambda_4
x+\lambda_6-\Big(x
\dfrac{v_3-u_3}{v_2-u_2}-\dfrac{u_2v_3-u_3v_2}{v_2-u_2}\Big)^2.$
Upon dividing the polynomial $\Phi(x, x^3+\lambda_4x+\lambda_6)$ by
the polynomial $(x-u_2)(x-v_2)$ we find
\begin{gather*}
\Phi(x,x^3+\lambda_4x+\lambda_6)\\ \qquad
=\Big(x+u_2+v_2-\Big(\frac{v_3-u_3}{v_2-u_2}\Big)^2\Big)(x^2-(u_2+v_2)x+v_2u_2)+\dots.
\end{gather*}
And, finally, we obtain the addition law of the elliptic groupoid in
the following form
\begin{gather*}
w_2=-(u_2+v_2)+h^2,\\
w_3=-\frac{1}{2}(u_3+v_3)+\frac{3}{2}(u_2+v_2)h-h^3,
\quad\text{where}\quad h=\Big(\frac{v_3-u_3}{v_2-u_2}\Big).
\end{gather*}

One may check directly that
$p_{\mathbb{C}^3}(A_3)=(\lambda_6,\lambda_4).$

\medskip Let $g=2.$ The family of curves $V$ is defined by the
polynomial
\begin{equation*}
f(x,y,\Lambda)=y^2-x^5-\lambda_4x^3-\lambda_6x^2-\lambda_8x-\lambda_{10}
\end{equation*}
In the coordinates
$(\Lambda_1,\Lambda_2)=((\lambda_{10},\lambda_8)^t,(\lambda_6,\lambda_4)^t)$
on $\mathbb{C}^4$ and $(P_{\even},P_{\odd},Z)$ on $\mathbb{C}^6,$
where $P_{\even}=(p_4,p_2)^t,$ $P_{\odd}=(p_5,p_3)^t$, and
$Z=(z_6,z_4)^t$, the mapping $p_{\mathbb{C}^6}$ is given by the
formula
\begin{gather*}(\Lambda_1,\Lambda_2)
=\bigg( \!\begin{pmatrix}p_5^2+p_3^2p_4-p_2p_4(p_2^2+p_4+z_4)-p_4(p_2p_4+z_6)\\
2p_3p_5+p_2p_3^2-(p_2^2+p_4)(p_2^2+p_4+z_4)-p_2(p_2p_4+z_6)
\end{pmatrix},Z\!\bigg)\end{gather*}
Let us write down the addition formulas for the points on the
groupoid  $\mathbb{C}^6$ over $\mathbb{C}^4$. Set
$A_1=((u_4,u_2)^t,(u_5,u_3)^t,(\lambda_4),\lambda_6)^t)$,
$A_2=((v_4,v_2)^t,(v_5,v_3)^t,(\lambda_4,\lambda_6)^t)$ and suppose
$p_{\mathbb{C}^6}(A_1)=p_{\mathbb{C}^6}(A_2)=((\lambda_{10},\lambda_8)^t,(\lambda_6,\lambda_4)^t)$.

Let  $A_3=A_1\star A_2,$
$A_3=((w_4,w_2)^t,(w_5,w_3)^t,(\lambda_6,\lambda_4)^t).$

We omit the calculation, which is carried out by the same scheme as
for $g=1$, and pass to the result. Set $h=h_1.$ We have
\begin{equation*}
h=-\frac{
\begin{vmatrix}
v_4-u_4&v_2v_4-u_2u_4\\
v_2-u_2&v_4+v_2^2-(u_4+u_2^2)
\end{vmatrix}}{
\begin{vmatrix}
v_4-u_4&v_5-u_5\\
v_2-u_2&v_3-u_3
\end{vmatrix}}.
\end{equation*}
To shorten the formulas it is convenient to employ the linear
differential operator
\begin{equation*}
\mathcal{L}=\frac{1}{2}\{(u_3-v_3)(\partial_{u_2}-\partial_{v_2})+
(u_5-v_5)(\partial_{u_4}-\partial_{v_4})\},
\end{equation*}
It is important to note that $\mathcal{L}$ adds unity to the weight,
$\deg\mathcal{L}=1,$ and that it is  \emph{tangent} to the singular
set where the addition is not defined:
\begin{equation*}
\mathcal{L}\{(u_2-v_2)(u_5-v_5)-(u_3-v_3)(u_4-v_4)\}=0.
\end{equation*}
Let  $h'=\mathcal{L}(h)$ and $h''=\mathcal{L}(h')$. Note, that
$\mathcal{L}(h'')=0$. Using this notation the addition formulas are
written down as follows
\begin{gather*}
w_2=\frac{1}{2}(u_2+v_2)+2h'+h^2,\\
w_3=\frac{1}{2}(u_3+v_3)+\frac{5}{4}(u_2+v_2)h+2h''+3h'h+h^3,\\
w_4=-\frac{1}{2}(u_4+v_4)-u_2v_2+
\frac{1}{8}(u_2+v_2)^2+(u_3+v_3)h\\ \qquad-\frac{1}{2}(u_2+v_2)(h'-h^2)-2h h'',\\
w_5=-\frac{1}{2}(u_5+v_5)-\frac{1}{2}(u_2u_3+v_2v_3)\\ \qquad-
   \big\{\frac{1}{8}(u_2+v_2)^2+v_2u_2+\frac{1}{2}(u_4+v_4)\big\}h
 +(u_3+v_3)(h'+h^2)  \\ \qquad-\frac{1}{2}(u_2+v_2)(h''-h h'-2h^3)-2(h'+h^2)h''.
\end{gather*}
\end{example}

\section{Addition theorems for hyperelliptic functions}
For each curve $V$ from the family \eqref{hypp} consider the Jacobi
variety  $\Jac(V)$. The set of all the Jacobi varieties is
\emph{the universal space $\mathsf{U}$ of the Jacobi varieties of
the genus $g$ hyperelliptic curves}.  The points of $\mathsf{U}$ are
pairs $(u,\Lambda),$ where the vector
 $u=(u_1,\dots,u_g)$ belongs to the Jacobi variety of the curve
 with parameters $\Lambda$. The mapping
${p}_{\mathsf{U}}:\mathsf{U}\to\mathbb{C}^{2g}$ that acts as
${p}_{\mathsf{U}}(u,\Lambda)=\Lambda$ makes $\mathsf{U}$  the space
over $\mathbb{C}^{2g}.$ There is a natural groupoid over
$\mathbb{C}^{2g}$ structure on  $\mathsf{U}$. Evidently, the
mappings
 $\mu((u,\Lambda),(v,\Lambda))=(u+v,\Lambda)$ and
$\inv(u,\Lambda)=(-u,\Lambda)$  satisfy the groupoid over
$\mathbb{C}^{2g}$ axioms.

\subsection{Addition theorems for the hyperelliptic
$\boldsymbol{\wp}$-functions} Let us define the mapping
$\pi:\mathsf{U}\to\mathbb{C}^{3g}$ over $\mathbb{C}^{2g}$ by putting
into correspondence a point $(u,\Lambda)\in \mathsf{U}$  and the
point $(\bwp(u),\bwp'(u)/2,\Lambda_2)\in\mathbb{C}^{3g}$, where
\[
\bwp(u)=(\wp_{g,j}(u))^t, \quad\bwp'(u)=(\wp_{g,g,j}(u))^t,\quad
\Lambda_{2}=(\lambda_{2(g-i+2)}),\;\;  i=1,\dots,g.\] Here
\[\wp_{i,j}(u)=-\partial_{u_i}\partial_{u_j}\log\sigma(u)\quad\text{and}\quad
\wp_{i,j,k}(u)=-\partial_{u_i}\partial_{u_j}\log\sigma(u) \] and
$\sigma(u)$ is the hyperelliptic sigma-function
\cite{ba97,ba07,bel97a,bel97b}.
\begin{theorem} The mapping  $\pi:\mathsf{U}\to\mathbb{C}^{3g}$
over $\mathbb{C}^{2g}$ is a birational isomorphism of groupoids:
\begin{equation*}
\pi(u+v,\Lambda)=\pi(u,\Lambda)\star\pi(v,\Lambda),\quad
\pi(-u,\Lambda)=\overline{\pi(u,\Lambda)}.
\end{equation*}\label{additio_2}
\end{theorem}
\begin{proof} First, by Abel theorem any triple of points
 $(u,v,w)\in (\Jac(V))^{3}$ that satisfies the condition
$u+v+w=0$  corresponds to the set of zeros $(x_i,y_i)$,
$i=1,\dots,3g$, of an entire rational function of order $3g$ on the
curve $V$. Namely, Let $X=(1,x,\dots,x^{g-1})^t$, then
\begin{equation}\label{x2uvw}
u=\sum_{i=1}^g\int_{\infty}^{x_i}X\,\frac{\mathrm{d}x}{2y},\quad
v=\sum_{i=1}^g\int_{\infty}^{x_{i+g}}X\,\frac{\mathrm{d}x}{2y},\quad
w=\sum_{i=1}^g\int_{\infty}^{x_{i+2g}}X\,\frac{\mathrm{d}x}{2y}.
\end{equation}
(For shortness, instead of indicating the end point of integration
explicitly, we give only the first coordinate.)

Second, for  the given value $u\in\Jac(V)$ the system of $g$
equations
\begin{equation*}
u-\sum_{i=1}^{g}\int_{\infty}^{x_i}X\frac{\mathrm{d}x}{2y}=0
\end{equation*}
with respect to the unknowns $(x_i,y_i)\in V$ is equivalent to the
system of algebraic equations
\begin{equation*}
x^g-\sum_{k=1}^g\wp_{g,k}(u)x^{k-1}=0,\quad
2y-\sum_{k=1}^g\wp_{g,g,k}(u)x^{k-1}=0,
\end{equation*}
the roots of which are the required points $(x_i,y_i)\in V$ (see,
for instance, \cite{ba97,bel97b}).

The combination of the two facts implies that the construction of
the preceding  sections provides the isomorphism.
\end{proof}
Above all note that $2g$ hyperelliptic functions\[
\bwp(u)=(\wp_{g,1}(u),\dots,\wp_{g,g}(u))^t\quad\text{and}\quad
\bwp'(u)=(\wp_{g,g,1}(u),\dots,\wp_{g,g,g}(u))^t\] form a basis of
the field of hyperelliptic Abelian functions, i.e., any function of
the field can be expressed as a rational function in $\bwp(u)$ and
$\bwp'(u)$. The assertion of Theorem \ref{additio_2} written down in
the coordinates of  $\mathbb{C}^{3g}$ takes the form of the addition
theorem for the basis functions $\bwp(u)$ and $\bwp'(u)$.
\begin{cor}\label{dura}
The basis hyperelliptic Abelian functions
\begin{equation*}\bwp(u)=(\wp_{g,1}(u),\dots, \wp_{g,g}(u))^t\quad\text{and}\quad
\bwp'(u)=(\wp_{g,g,1}(u),\dots,\wp_{g,g,g}(u))^t\end{equation*}
respect the addition law
\begin{equation*}
(\bwp(u+v),\bwp'(u+v)/2,\Lambda_2)=
(\bwp(u),\bwp'(u)/2,\Lambda_2)\star(\bwp(v),\bwp'(v)/2,\Lambda_2),
\end{equation*}
the formula of which is given in Theorem \ref{zakon}.
\end{cor}
Thus we have obtained a solution the problem to construct  an
explicit and effectively computable formula of the addition law in
the fields  of hyperelliptic Abelian functions.

\subsection{Addition theorems for the hyperelliptic   $\zeta$-functions}
One has  $g$ functions $\zeta_{i}(u)=\partial_{u_i}\log\sigma(u)$
and  the functions are not Abelian. However, by an application of
Abel theorem for the second kind integrals (see \cite{ba97}) one
obtains the addition theorems for $\zeta$-functions as well. On one
hand, any $\zeta$-function can be represented as the sum of $g$
second kind integrals and an Abelian function. On the other hand, an
Abelian sum of the second kind integrals with the end points at the
set of zeros of an entire rational function $R(x,y)$ is expressed
rationally in terms of the coefficients of $R(x,y)$. We employ the
function \eqref{R_explicit} computed in the variables indicated in
Corollary \ref{dura}.
\begin{theorem}\label{zeta_g} Let $(u,v,w)\in(\Jac(V))^{3}$ and $u+v+w=0.$
Then
\begin{gather*}
\zeta_g(u)+\zeta_g(v)+\zeta_g(w)=-h_1,
\end{gather*}
where $h_1$ is the rational function in $\bwp(u),\bwp'(u)$ and
$\bwp(v),\bwp'(v)$ equal to the coefficient of the monomial of the
weight $3g-1$ in the function \eqref{R_explicit} computed in the
variables indicated in Corollary \ref{dura}.
\end{theorem}
\begin{proof}
We have the identity (see \cite{ba97},\cite[p.~41]{bel97b})
\begin{gather*}\zeta_g(u)+\sum_{i=1}^g\int_{\infty}^{x_i}x^g\,\frac{\mathrm{d}x}{2y}=0,\quad
\zeta_g(v)+\sum_{i=1}^g\int_{\infty}^{x_{i+g}}x^g\,\frac{\mathrm{d}x}{2y}=0,\\\qquad
\zeta_g(w)+\sum_{i=1}^g\int_{\infty}^{x_{i+2g}}x^g\,\frac{\mathrm{d}x}{2y}=0.
\end{gather*}
Suppose that the closed path $\gamma$ encloses all zeros
$(x_1,y_1),\dots,(x_{3g},y_{3g})$ of the function $R_{3g}(x,y)$.
Then we have
\begin{gather*}
\sum_{k=1}^{3g}\int_{\infty}^{x_k}x^g\,\frac{\mathrm{d}x}{2y}=\frac{1}{2\pi\imath}
\oint_{\gamma}\mathrm{d}\big(\log
R_{3g}(x,y)\big)\int_{\infty}^{x}x^g\,\frac{\mathrm{d}x}{2y}.
\end{gather*}
 Because $\mathrm{d}\log
R_{3g}(x,y)/\mathrm{d} x$ is a rational function on the curve and,
hence, a uniform function, the total residue of $\mathrm{d}\big(\log
R_{3g}(x,y)\big)\int_{\infty}^{x}x^g\mathrm{d}x/(2y)$ on the Riemann
surface of the curve $V$ is zero. To write down this fact explicitly
consider the parametrization
\begin{equation*}
(x(\xi),y(\xi))=(\xi^{-2},\xi^{-2g-1}\rho(\xi)), \qquad
\rho(\xi)=1+\frac{\lambda_4}{2}\xi^4+\frac{\lambda_6}{2}\xi^6+O(\xi^8),
\end{equation*}
of the curve $V$ near the point at infinity and denote
$R_{3g}(\xi)=R_{3g}(x(\xi),y(\xi))$. We obtain
\begin{equation*}-
\mathrm{Res}_{\xi}\bigg[\frac{
R_{3g}'(\xi)}{R_{3g}(\xi)}\int_{\infty}^{x(\xi)}x^g\,\frac{\mathrm{d}x}{2y}\bigg]+\sum_{i=1}^{3g}
\mathrm{Res}_{x=x_i}\!\bigg[\mathrm{d}\big(\log
R_{3g}(x,y)\big)\int_{\infty}^{x}x^g\,\frac{\mathrm{d}x}{2y}\bigg]\!=0,
\end{equation*}
which is in fact a particular case of Abel theorem. Thus, the final
expression is
\[
\zeta_g(u)+\zeta_g(v)+\zeta_g(w)=-\mathrm{Res}_{\xi}\bigg[\frac{
R_{3g}'(\xi)}{R_{3g}(\xi)}\int_{\infty}^{x(\xi)}x^g\,\frac{\mathrm{d}x}{2y}\bigg].
\]
It remains  to use the expansions
\begin{gather*}
\int_{\infty}^{x(\xi)}x^g\,\frac{\mathrm{d}x}{2y}=\frac{1}{\xi}
+\frac{\lambda_4}{6}\xi^3+O(\xi^5),\\ R_{3g}(\xi)=
\xi^{-3g}(1+h_1\xi+h_2\xi^2+h_3\xi^3+O(\xi^4)).
\end{gather*}
to compute the residue.
\end{proof}
A similar argument leads from the identity (see
\cite{ba97},\cite[p.~41]{bel97b})
\begin{equation*}\zeta_{g-1}(u)+\sum_{i=1}^g\int_{\infty}^{x_i}
(3x^{g+1}+\lambda_4
x^{g-1})\dfrac{\mathrm{d}x}{2y}=\frac{1}{2}\wp_{g,g,g}(u),
\end{equation*} to the following assertion.
\begin{theorem}In the conditions of Theorem \ref{zeta_g}  we have
\begin{gather*}
\zeta_{g-1}(u)+\zeta_{g-1}(v)+\zeta_{g-1}(w)\\\qquad-\frac{1}{2}(
\wp_{g,g,g}(u)+\wp_{g,g,g}(v)+\wp_{g,g,g}(w))=-h_1^3+3h_1h_2-3h_3,
\end{gather*}
\label{zeta_g-1} where $h_2$ and $h_3$ are the  coefficients of the
monomials of weight $3g-2$ and $3g-3$ in the function indicated in
Theorem \ref{zeta_g}.
\end{theorem}
\begin{example}\label{ex:g1} Let $g=1$. The function $R_3(x,y)$ has the form $y+h_1
x+h_3$, where $2h_1=(\wp'(u)-\wp'(v))/(\wp(u)-\wp(v))$,  cf. Example
\ref{genus1.1}. Thus, Theorem \ref{zeta_g} gives the classic formula
\begin{equation}
\zeta(u)+\zeta(v)-\zeta(u+v)=-\frac{1}{2}\bigg(\frac{\wp'(u)-\wp'(v)}{\wp(u)-\wp(v)}\bigg),\quad
u+v+w=0.\label{FS_1}
\end{equation}
As $h_2=0$ and $2h_3=(\wp'(v)\wp(u)-\wp'(u)\wp(v))/(\wp(u)-\wp(v))$,
cf. Example \ref{genus1.1}, Theorem \ref{zeta_g-1}  yields the
relation
\begin{equation*}
-\wp'(u)-\wp'(v)+\wp'(u+v)=-\frac{1}{4}\bigg(\frac{\wp'(u)
-\wp'(v)}{\wp(u)-\wp(v)}\bigg)^3-3\frac{\wp'(v)\wp(u)-\wp'(u)\wp(v)}{\wp(u)-\wp(v)},
\end{equation*}
which is the addition formula for Weierstrass $\wp'$-function.
\end{example}
The fact below follows directly from Lemma \ref{even_part}.
\begin{lemma}\label{p_gg} $
\wp_{g,g}(u)+\wp_{g,g}(v)+\wp_{g,g}(u+v)=h_1^2-2h_2.$
\end{lemma}
Combining Lemma \ref{p_gg} with Theorem \ref{zeta_g} we find
\begin{equation}\label{FS_g}
\big(\zeta_g(u)+\zeta_g(v)+\zeta_g(w)\big)^2=
\wp_{g,g}(u)+\wp_{g,g}(v)+\wp_{g,g}(u+v)+2h_2
\end{equation}
In the case $g=1$  due to the fact that $h_2=0$  formula
\eqref{FS_g} gives the famous relation
 \begin{equation*}
(\zeta(u)+\zeta(v)-\zeta(u+v))^2=\wp(u)+\wp(v)+\wp(u+v).
 \end{equation*}
discovered by Frobenius and Stickelberger.

\begin{example} Let us pass to the case $g=2$, we have $R_6(x,y)=x^2+h_1 y+h_2 x^2+h_4
 x+h_6$. Note that $h_3=0$.  The coefficient  $h_1$ is expressed as
follows, cf. Example \ref{genus1.1},
\begin{equation*}
h_1=-2\frac{
\begin{vmatrix}
\wp_{2,1}(v)-\wp_{2,1}(u)&\wp_{2,2}(u)\wp_{2,1}(v)-\wp_{2,2}(v)\wp_{2,1}(u)\\
\wp_{2,2}(v)-\wp_{2,2}(u)&\wp_{2,1}(v)-\wp_{2,1}(u)
\end{vmatrix}}{
\begin{vmatrix}
\wp_{2,1}(v)-\wp_{2,1}(u)&\wp_{2,2,1}(v)-\wp_{2,2,1}(u)\\
\wp_{2,2}(v)-\wp_{2,2}(u)&\wp_{2,2,2}(v)-\wp_{2,2,2}(u)
\end{vmatrix}}.
 \end{equation*}
And the coefficient $h_2$, respectively,
\begin{equation*}
h_2=\frac{
\begin{vmatrix}
\wp_{2,2}(v)\wp_{2,1}(v)-\wp_{2,2}(u)\wp_{2,1}(u)&\wp_{2,2,1}(v)-\wp_{2,2,1}(u)\\
\wp_{2,1}(v)+\wp_{2,2}(v)^2-\wp_{2,1}(u)-\wp_{2,2}(u)^2&\wp_{2,2,2}(v)-\wp_{2,2,2}(u)&
\end{vmatrix}}{
\begin{vmatrix}
\wp_{2,1}(v)-\wp_{2,1}(u)&\wp_{2,2,1}(v)-\wp_{2,2,1}(u)\\
\wp_{2,2}(v)-\wp_{2,2}(u)&\wp_{2,2,2}(v)-\wp_{2,2,2}(u)
\end{vmatrix}}.
 \end{equation*}
We come to the relations
\begin{gather*}
\zeta_2(u)+\zeta_2(v)-\zeta_2(u+v)=-h_1,\\
\wp_{2,2}(u)+\wp_{2,2}(v)+\wp_{2,2}(u+v)=h_1^2-2h_2,\\
\zeta_{1}(u)+\zeta_{1}(v)-\zeta_{1}(u+v)\\
\qquad-\frac{1}{2}(
\wp_{2,2,2}(u)+\wp_{2,2,2}(v)-\wp_{2,2,2}(u+v))=-h_1^3+3h_1h_2.
\end{gather*}
Hence, by eliminating $h_1$ and $h_2$, we obtain the identity
\begin{equation}
2\mathfrak{z}_{1}-\mathfrak{p}_{2,2,2}-3\mathfrak{p}_{2,2}\mathfrak{z}_2+\mathfrak{z}_2^3=0,\label{z-p}
\end{equation}
where $\mathfrak{z}_{i}=\zeta_i(u)+\zeta_i(v)+\zeta_i(w)$ and
$\mathfrak{p}_{i,j,\dots}=\wp_{i,j,\dots}(u)+\wp_{i,j,\dots}(v)+\wp_{i,j,\dots}(w)$,
provided $u+v+w=0.$
\end{example}
\subsection{Trilinear addition theorems for the hyperelliptic $\sigma$-functions}
Formula \eqref{z-p} leads to an important corollary.
\begin{theorem} The genus $2$ sigma-function respects the trilinear addition law
\begin{gather*}
\big[2D_1+ D_2^3\big]\sigma(u)\sigma(v)\sigma(w)\big|_{u+v+w=0}=0,
\end{gather*}
where
$D_j=\partial_{u_j}+\partial_{v_j}+\partial_{w_j}$.\label{trilinea}
\end{theorem}
\begin{proof}
Let us multiply the left hand side of \eqref{z-p} by the product
$\sigma(u)\sigma(v)\sigma(w)$, then \eqref{z-p} becomes the
trilinear relation
\begin{gather*}
\big[2(\partial_{u_1}+\partial_{v_1}+\partial_{w_1})+
(\partial_{u_2}+\partial_{v_2}+\partial_{w_2})^3\big]\sigma(u)\sigma(v)\sigma(w)\big|_{u+v+w=0}=0,
\end{gather*}
which is satisfied by the genus $2$ sigma-function.
\end{proof}
It is important to notice that the elliptic identity \eqref{FS_1} is
equivalent to the trilinear addition law
\begin{gather*}
\big[(\partial_{u}+\partial_{v}+\partial_{w})^2\big]\sigma(u)\sigma(v)\sigma(w)\big|_{u+v+w=0}=0,
\end{gather*}
which is satisfied by Weierstrass sigma-function.
 Let us denote
$D=(\partial_{u}+\partial_{v}+\partial_{w})$ and
$\psi=\sigma(u)\sigma(v)\sigma(w)$. The functions
\[
(D+h_1)\psi,\quad (D^3+6h_3)\psi, \quad (D^4-6\lambda_4)\psi,  \quad
(D^5+18\lambda_4 D)\psi,\quad (D^6-6^3\lambda_6)\psi,
\]
where $h_1$ and $h_2$ are given in Example \ref{ex:g1}, vanish on
the plane $u+v+w=0$. Moreover, one can show that for any $k>3$ there
exist unique polynomials $q_0, q_1, q_3\in
\mathbb{Q}[\lambda_4,\lambda_6]$ such that
\[
(D^k+q_3D^3+q_1 D+q_0)\psi\big|_{u+v+w=0}=0,
\]
and at least one of the polynomials $q_0,q_1,q_3$ is nontrivial.

For the hyperelliptic sigma-function of an arbitrary genus $g$ we
propose the following hypothesis. Let
$\mathscr{P}=\mathbb{Q}[\Lambda]$. Consider the ring
$\mathscr{Q}=\mathscr{P}[D_1,\dots,D_g]$ as a graded ring of linear
differential operators. We conjecture that there exists a collection
of $3g$ linear operators $Q_i\in\mathscr{Q},$ $\deg Q_i=i,$ where
$i=1,\dots, 3g$, such that
\[
\Big\{\sum_{i=0}^{3g}Q_i\xi^{3g-i}+R_{3g}(\xi^2,\xi^{2g+1})\Big\}\sigma(u)\sigma(v)\sigma(w)\big|_{u+v+w=0}=0,
\]
where $Q_0=1$ and $R_{3g}(x,y)$ is the function \eqref{R_explicit}
computed in the variables indicated in Corollary \ref{dura}. Thus,
$g$ operators $Q_{g+2i-1},$ $i=1,\dots,g$ define the trilinear
relations
\[
Q_{g+2i-1}\sigma(u)\sigma(v)\sigma(w)\big|_{u+v+w=0}=0,\quad
i=1,\dots,g.
\]
Note, that the assertions of Theorem \ref{zeta_g}, Lemma \ref{p_gg},
and Theorem \ref{zeta_g-1} imply the relations
\begin{gather*}
(D_g+h_1)\sigma(u)\sigma(v)\sigma(w)\big|_{u+v+w=0}=0,\\
(D_g^2-2h_2)\sigma(u)\sigma(v)\sigma(w)\big|_{u+v+w=0}=0,\\
(2D_{g-1}+D_g^3+6h_3)\sigma(u)\sigma(v)\sigma(w)\big|_{u+v+w=0}=0.
\end{gather*}

We shall return to the problem of explicit description of the
trilinear addition theorems for the hyperelliptic sigma-function in
our future publications.

\end{document}